\documentstyle[aps]{revtex}

\def\be{\begin{equation}}
\def\ee{\end{equation}}
\def\bea{\begin{eqnarray}}
\def\eea{\end{eqnarray}}
\def\ben{\begin{eqnarray*}}
\def\een{\end{eqnarray*}}
\def\pl{\partial}
\def\th{\theta}

\begin{document}

\title{The most general axially symmetric electrovac spacetime adimitting 
separable equations of motion}

\author{Naresh Dadhich$^a$ ,\,  Z. Ya. Turakulov$^{b}$}
\address{Inter-University Centre for Astronomy and Astrophysics,
Post Bag 4, Ganeshkhind, Pune~411~007, India.}


\maketitle
\begin{abstract}
 We obtain the most general solution of the Einstein electro - vacuum 
equation for the stationary axially symmetric spacetime in which the 
Hamilton-Jacobi and Klein - Gordon equations are separable. The most 
remarkable feature of the solution is its invariance under the 
duality transformation involving mass and  NUT parameter, and the radial 
and angle coordinates. It is the general solution for a rotating 
(gravitational dyon) particle 
which is endowed with both gravoelectric and gravomagnetic charges, and there 
exists a duality transformation from one to the other. It also happens to be 
a transform of the Kerr - NUT solution. Like the Kerr family, it is also 
possible to make this solution radiating which asymptotically conforms 
to the Vaidya null radiation.
\end{abstract}


\vspace{0.5cm}


\vspace{0.5cm}

\section{Introduction}

 In spherical symmetry, vacuum spacetime is characterized by the Schwarzschild 
solution because the Einstein vacuum field equation can admit no other 
solution. The situation is however different for axial symmetry. First we 
have to assume stationarity. For the stationary axially symmetric spacetime, 
the Kerr solution is unique only under the assumptions of existence of 
regular horizon and asymptotic flatness. There exist asymptotically flat 
generalization of the Kerr solution in Tomimatsu - Sato series solutions 
[1,2] and asymptotically non-flat in the Kerr - NUT solution [3]. 

 In this paper, we shall obtain the most general solution of the Einstein 
electro-vacuum equation  
for the stationary axially symmetric spacetime metric in which motion is 
integrable. That is, the Hamilton - Jacobi (HJ) equation for particle motion 
and the Klein - Gordon (KG) equation for scalar field propagation are 
separable. We shall henceforth term it as integrable spacetime (metric). The 
separability of the equations of motion was first considered long back by 
Carter [4]. In particular, the separability of HJ and KG equations was 
employed by one of us (ZYT) for obtaining the 
Kerr solution [5]. Integrability of motion is the primary requirement for 
physical understanding of spacetime. Under this  physically motivated 
assumption, we shall seek the most general electrovac solution and show that 
it happens to be a transform of the Kerr - NUT solution with  electric 
charge.  The most general stationary axially symmetric 
electrovac spacetime with separable equations of motion is therefore 
characterized by the four parameters. 
They are the usual mass, spin and electric charge, and the NUT parameter. This 
spacetime is however not asymptotically flat, and the departure from 
it is the measure of the NUT parameter.

 We first implement  separability of the two equations in the most general 
stationary axially 
symmetric spacetime metric. This would lead to determining form and character 
of the metric coefficients. For  this metric, we shall now solve the Einstein 
equation, which would readily give the most general solution for the 
electrovac spacetime. The unique solution so obtained could be transformed to
the Kerr - NUT solution. If we now impose asymptotic flatness, it would 
reduce to the Kerr 
family. This is therefore a very direct and straight forward way of 
establishing the uniqueness of the Kerr family as well. This happens because 
our assumptions are more general than those of the Kerr family. It is 
well-known that motion is indeed integrable in the Kerr metric. Note that 
integrability of motion does imply existence of regular horizon but not 
asymptotic flatness. That is why the general family is the 
asymptotically non flat Kerr - NUT family [3,6] which reduces to the Kerr 
family when asymptotic flatness is imposed. 

 One of the remarkable features of our derivation of the general solution is 
that it brings forth the duality between the mass $M$, and the NUT parameter, 
$l$. The solution is invariant under the duality transformation, 
$M\leftrightarrow il, \, r\leftrightarrow ia\lambda$, where $\lambda$ is an 
angle coordinate and $a$ is the rotation parameter. This is the duality 
between gravoelectric ($M$) and gravomagnetic ($l$) charges. This is therefore
the most general solution for a localized source which is endowed with both 
gravitational electric and magnetic charges.  When $l=0$, it 
is the Kerr spacetime. On the other hand when $M=0$, it is a vacuum spacetime 
which is dual to the Kerr spacetime [7]. One can go from one to the other by  
$M \leftrightarrow il$ and $r \leftrightarrow ia\lambda$;i.e. the Kerr 
solution goes over to the dual solution and the vice - versa. Though Kerr - 
NUT solution has been 
known for a long time, yet the dual to Kerr solution has not been considered 
earlier than [7]. Here the duality springs up quite naturally and obviously. 

 There exist a number of radiating generalizations of the Kerr solution 
[8-10]. One of them relies on making mass function of the retarded Eddington 
time, and then the spacetime has the Ricci scalar $R=0$, and it asymptotically
 conforms to the Vaidya null radiation [8,9]. The same prescription could be 
effected for the Kerr - NUT solution as well to make it radiating with 
vanishing Ricci scalar and asymptotically conforming to the Vaidya null 
radiation.

 The paper is organized as follows: In the next section, we shall derive the 
stationary axially symmetric integrable metric in which motion is integrable. 
In Sec. III, we shall solve 
the Einstein equation to obtain the most general solution and establish the 
uniqueness of the Kerr - NUT family. Sec. IV will be devoted to the 
consideration of the dual Kerr solution and the radiating generalization of 
the Kerr - NUT solution. We shall conclude with a discussion.

\section{The integrable metric}

 Any coordinate system $\{x^i\}$ in the spacetime specifies natural vector
$\{\pl_i\}$ and covector $\{dx^i\}$ frames at each point. The metric in the 
covariant form is defined by the scalar product of the vectors, 
$g_{ij}= <\pl_i,\pl_j>$ while in the contravariant form by the scalar product 
of the covectors, 
$g^{ij}= <dx^i,dx^j>$. The stationarity and axial symmetric character 
of spacetime is characterized by existence of the two familiar Killing 
vectors, $\pl_t$ and $\pl_\varphi$. The intersections of $t=const.$
and $\varphi=const.$ are 2-dimensional surfaces which can be endowed with
an orthogonal coordinate system $(u,v)$. 

 We introduce the coordinates $(t,u,v,\varphi)$ and write the most general
 stationary and axially symmetric metric in the contravariant $g^{ij}$ form as 
follows:  
\bea 
<d\varphi,d\varphi>=-A,\ <dt,dt>=B,\ <dt,d\varphi>=
\Gamma,\\ \nonumber <du,du>=-\bar U(u)^2/\Sigma, <dv,dv>=-\bar V(v)^2/\Sigma.
\eea 

 The contravariant form is chosen because this is how the metric appears in 
HJ equation, that is what we consider next. HJ equation for geodesics of the 
above metric is given by 
\be 
-A(\pl_\varphi S)^2+B(\pl_t S)^2+2\Gamma \pl_t S\pl_\varphi S-\Sigma^{-1}
\left((\pl_u S)^2+(\pl_v S)^2\right) = m^2
\ee
where $m$ is the rest mass of the particle. As usual, the separability of the 
equation leads to 
\be
 S= Et + L\varphi + f(u) + g(v)
\ee 
where $E$ and $L$ are the two constants of motion corresponding to the two 
Killing vectors, $\pl_t$ and $\pl_\varphi$. 

 Substituting Eq. (3) into Eq. (2) would readily lead to 
\bea 
A \,\Sigma &=& \nonumber -U_{33}(u)+V_{33}(v),\ B \,\Sigma=U_{00}(u)-V_{00}(v),\\  
\Sigma &=& F(u)+G(v), \Gamma \,\Sigma = U_{03}(u)+V_{03}(v).
\eea
 Here $U_{ab}, \ V_{ab}, \ a,b=0,3$ and $ F, \ G$ are arbitrary functions of 
their arguments. The determinant of the metric is given by
\be
-g=\Sigma^2(AB+\Gamma^2)^{-1}(\bar U\bar V)^{-2}.
\ee 
 Let us now turn to KG equation, which reads as
\ben
\frac{1}{\sqrt{-g}}(\sqrt{-g}g^{ab}\phi_{,b})_{,a}=0
\een
where comma denotes the partial derivative. Now separability of this equation 
leads to the separability of the block determinant, $AB+\Gamma^2$. That means
\ben
\frac{(V_{33}-U_{33})(U_{00}-V_{00}) + (U_{03}+V_{03})^2}{(F+G)^2} 
\een
must be separable; i.e. of the form $(U(u)V(v))^{-2}$, say. We could hence 
write
\be
(U_{03}^2-U_{00}U_{33})+(V_{03}^2-V_{00}V_{33})+(U_{00}V_{33}+
2U_{03}V_{03}+V_{00}U_{33}) = \nonumber  [U(u)V(v)]^{-2}(F+G)^2.
\ee

 Note that $\Sigma$ should have dimension of length square, which would mean 
the functions $F(u),G(v)$ should be of this dimension. For $F(u)$, it simply 
means that it is quadratic in $u$, while $G(v)$ must be given this dimension  
by introducing a constant, say $a$ of the length dimension. Evidently $A, B$ 
and $\Gamma$ would respectively be of dimension, $(L^{-2},\,L^0, \,L^{-1})$, 
and similarly $U_{00}, \,V_{00}$ of $L^2$, $U_{33}, \,V_{33}$ of $L^0$, and 
$U_{03}, \ V_{03}$ of $L$.  

 Let us now write $UU_0=F, UU_3=a,VV_0=G/a,VV_3=1$. This entails no loss of 
generality as we have simply written the four functions $(U_0,U_3,V_0,V_3)$ 
in terms of the arbitrary four functions $(U,V,F,G)$. With this, the right 
hand side becomes $(U_0V_3+U_3V_0)^2$. All the functions $U_{ab},V_{ab}$ could 
also be written as quadratic functions of $(U_0,U_3,V_0,V_3)$. A straight 
forward calculation based on Eq. (6) would then lead to the unique 
identification, 
$U_{00}=U_0^2, U_{03}=U_0U_3, U_{33}=U_3^2$, and similarly for $V_{ab}$. The 
first two terms on the left would vanish and the third term will reduce to 
$(U_0V_3+U_3V_0)^2$ conforming with the right hand side. The metric functions 
will now read as
\bea
A \,\Sigma &=& -\frac{a^2}{U^2}+\frac{1}{V^2},\ B \,\Sigma=\frac{F^2}{U^2}-
\frac{G^2}{a^2V^2}, \nonumber \\ 
\Gamma \,\Sigma &=& \frac{aF}{U^2}+\frac{G}{aV^2}, \nonumber \\
 \Sigma &=& F+G.
\eea
We have thus arrived at the integrable metric in which HJ and KG equations 
are separable, and it is given by
\ben
<dt,dt>=\frac{F^2U^{-2}-G^2(aV)^{-2}}{F+G},\ <dt,d\varphi>=
a\frac{U^{-2}F+(aV)^{-2}G}{F+G},\\ \nonumber <d\varphi,d\varphi>=
\frac{a^2U^{-2}-V^{-2}}{F+G},\ <dr,dr>=-\frac{U^2}{F+G},\ 
<d\lambda,d\lambda>=-\frac{V^2}{F+G}.
\een 
that would read in the covariant form as follows:
\be
ds^2=\frac{U^2-a^2V^2}{F+G}dt^2+2a\frac{FV^2+a^{-2}GU^2}{F+G}dtd\varphi-(F+G)
(\frac{dr^2}{U^2}+\frac{d\lambda^2}{V^2})
-\frac{F^2V^2-a^{-2}G^2U^2}{F+G}d\varphi^2
\ee
where we have redefined the radial, $du/\bar U(u)=dr/U(r)$ and the angle, 
$dv/\bar V(v)=d\lambda/V(\lambda)$ coordinates. This is the integrable metric 
satisfying the condition of separability of HJ and KG equations. 
Alternatively, the 
separability of KG equation follows from the separability of HJ equation and 
the Einstein vacuum ($T_{r\theta}=0$) equation [11].

 Note that separability of the block determinant is 
also required for existence of horizon. The horizon is defined when the 
timelike corotating vector, $\pl_t+\omega\pl_\varphi,\ \omega=-g_{t\varphi}/
g_{\varphi\varphi}$, turns null on the coordinate surface, $r=const$. It is 
easy to see that norm of this vector is proportional to the block determinant,
 and hence its separability would incorporate existence of horizon. Our 
integrable metric thus shares the property of existence of horizon with the 
Kerr family but it need not be asymptotically flat. This opens up the window 
for non asymptotically flat generalization of the Kerr family.

 Alternatively, if we consider the null geodesics ($m=0$), Eq.(2) would reduce 
to the following two equations,
\ben
 U_{00}E^2+2U_{03}EL+U_{33}L^2-f'^2=C\\ \nonumber 
V_{00}E^2-2V_{03}EL+V_{33}L^2+g'^2=C
\een 
where $C$ is an arbitrary constant. It turns out that separability of KG 
equation could be traded off with the requirement that the quadratic 
expressions above are perfect squares. 

 It is known that black hole vacuum spacetimes admit event horizon which is a
null surface with null geodesics as its generators. That means no null 
geodesic with its tangent vector lying on the horizon can leave it. It would 
be a coordinate surface, $r=const.$ on which $U=0$. Consider the null geodetic 
flow specified by the three integrals of motion $E,\ L$ and $C$. The flow is 
defined in the region in which the momenta, $p_u=f'$ and $p_v=g'$:
\ben
f'^2=(U_0^2E^2+2U_0U_3EL+U_3^2L^2)-C=(U_0E+U_3L)^2-C\\
g'^2=C-(V_0^2E^2-2V_0V_3EL+V_3^2L^2)=C-(V_0E-V_3L)^2
\een
take real values. The integral of motion $C$ cannot be taken zero because in
that case $g'$ becomes imaginary everywhere. Clearly, any change of this
constant changes both components of the flow. In terms of $\{r,\lambda\}$
variables the components take the form
$$
p_r=\sqrt{(FE+aL)^2-CU^2},\ p_\lambda=\sqrt{CV^2-(GE-aL)^2}.
$$
At the horizon both $p_r,\ p_r'$ must vanish ensuring that null geodesics 
lying on it cannot leave it. The horizon is defined by $U(r)=0$ and hence 
$EF+aL=0$ at the horizon. That means for existence of horizon, the quadratic 
expressions above must be perfect squares. This would again lead to the 
integrable metric (8). Thus the requirement of existence of horizon is 
equivalent to the separability of the block determinant which is required for 
separability of KG equation.  

\section{The general solution}

 The orthonormal tetrad frame for the metric (8) is given by
\ben
\nu^0=U\sigma(dt+a^{-1}Gd\varphi),\ \nu^1=(U\sigma)^{-1}dr, \\ \nonumber
\nu^2=(V\sigma)^{-1}d\lambda,\ \nu^3=V\sigma(-adt+Fd\varphi)\\
\een
where
$$
\sigma=(F+G)^{-1/2}.
$$

 We obtain the non-zero Ricci components in this tetrad frame and they are 
given by
\ben
R_{03}&=&-\frac{1}{2}(aUV\sigma^4)(F''-a^{-2}G''),\\
R_{00}&=&-\frac{1}{2}(U^2-a^2V^2)(F'^2+a^{-2}G'^2)\sigma^6-
(\frac{U^2}{2})''\sigma^2+\\ \nonumber&+&(U^2+UU'F'-a^2V^2-VV'G')\sigma^4,\\
R_{33}&=&-\frac{1}{2}(U^2-a^2V^2)(F'^2+a^{-2}G'^2)\sigma^6+
(\frac{V^2}{2})''\sigma^2+\\ \nonumber&+&(U^2+UU'F'-a^2V^2-VV'G')\sigma^4,\\
R_{11}&=&(\frac{U^2}{2})''\sigma^2+(U^2-UU'F'-a^2V^2+VV'G')\sigma^4,\\
R_{22}&=&-\frac{1}{2}U^2(F'^2+a^{-2}G'^2)\sigma^6+
(\frac{V^2}{2})''\sigma^2+\\ \nonumber&+&(U^2+UU'F'+a^2V^2-VV'G')\sigma^4.
\een

 Let us first implement the electrovac Einstein equation as vacuum would be 
contained in it. That is, $R_{00}=-R_{11}=R_{22}=R_{33}$ and the rest being 
zero. Now $R_{03}=0$ gives
\be
F''=a^{-2}G''=const.
\ee
and $R_{00}=R_{33}$ leads to
\be
(U^2)''= -(V^2)''=const.
\ee
Using them in $R_{00}=-R_{11}$ yields
\be
F'^2+a^{-2}G'^2=4(F+G)=4\sigma^{-2}.
\ee
Then a remarkable reduction occurs and all the non-zero $R_{ab}$ would 
collapse to
\be
R_{00}=R_{33}=-R_{11}=R_{22}=\frac{-U^2+UU'F'+a^2V^2-VV'G'-F-G}{(F+G)^2}.
\ee
 Note that Eq. (11) would require that the constants in Eqs. (9) and (10) 
cannot be different. Further, linear term in the argument for $F$ and $G$ 
can be transformed away, and hence they will only contain the quadratic term. 
Then from Eqs (9-11), we thus obtain the general electrovac solution given by
\be
U^2=r^2-2Mr+p, \ V^2=-\lambda^2+2N\lambda+q, \ F=r^2+a^2, \ G=a^2(\lambda^2-1) 
\ee
and then Eq.(12) would read as,
\be
R_{00}=-R_{11}=R_{22}=R_{33}=\frac{-p+a^2q}{(F+G)^2} = \frac{Q^2}{(F+G^2)}
\ee
which would reduce to vacuum when
\be
p=a^2q, \ Q=0.
\ee
It is clear that the constant $Q^2=-p+a^2q$ would denote the electric charge 
on the source particle. The parameter $a$ having the dimension of length was 
introduced while implementing the condition of separability of HJ equation. 
In the Kerr solution, it represents the usual rotation parameter. The 
remaining four parameters arise as constants of integration, of which $M$ and 
$Q$ represent mass and electric charge. The other two are dimensionless. It 
would turn out that the parameter $q$ could be transformed away while $l=aN$ 
would denote the NUT parameter. 

 We have thus obtained the general solution, hence unique, of the integrable 
metric given by (8), and it is given by
\be
ds^2 = \Lambda(dt + \alpha d\varphi)^2 - (\Lambda)^{-1}\left[(U^2 - a^2V^2)
\left(\frac{dr^2}{U^2} + \frac{d\lambda^2}{V^2}\right) 
+ U^2V^2 d\varphi^2\right] 
\ee
where 
\bea
\Lambda &=& \frac{U^2 - a^2V^2}{F + G}, \nonumber \\
\alpha &=& a\frac{(F - U^2)V^2 + (Ga^{-2} + V^2)U^2}{U^2 - a^2V^2},\nonumber \\
 F &=& r^2 + a^2, \ G = a^2(\lambda^2 - 1), \nonumber \\
 U^2&=& r^2-2Mr+p, \ V^2=-\lambda^2 + 2N\lambda + q.
\eea

 The most remarkable feature of this vacuum metric with $Q = 0$ is that it is 
invariant under the transformation $M \leftrightarrow il, \, r\leftrightarrow 
ia\lambda$ where $l = aN$. For the elctrovac case, we should in addition have 
$p \leftrightarrow aq^2$ which would imply $Q^2 \leftrightarrow -Q^2$. It is 
easy to verify that $U^2 \leftrightarrow a^2V^2, \, 
F \leftrightarrow -G$. This indicates the duality between mass (gravoelectric 
charge) and the NUT parameter (gravomagnetic charge [12]). In this duality 
transformation the parameter $a$ plays the crucial role. The transformation 
is defined only when it is non zero. This is therefore the most 
general solution for a localized source having both gravitational electric 
and magnetic charges. It is the magnetic charge that requires the solution 
to be asymptotic non flat. By the duality transformation, gravoelectric 
particle (Kerr) could be transformed into gravomagnetic (dual Kerr) particle.

 When $a = 0$, it reduces to the Reissner - Nordstr${\ddot o}$m  (RN) 
solution of a charged black hole with a deficit angle. The dimensionless 
parameters attain non trivial physical meaning only in the presence of $a$. 
It turns out that $l = aN$ can represent the NUT parameter while the other 
one is redundant 
which could be transformed away. We shall now transform this solution to the 
Kerr - NUT form [3], which has only four  
parameters, $M,\ a, \ Q, \ l$ representing the NUT generalization of Kerr  
family [3]. The transformation that does this, proceeds as follows: 
 \ben
q &=& (b^2 - l^2)/a^2, \ N = l/a, \ a^2 = d^2 + b^2 + l^2,\\ 
\varphi &=& (a/b)\bar\varphi, \ t = \bar t + (d^2/b)\bar \varphi, \\
\bar\alpha &=& (a \alpha +d^2)/b.
\een
 This would lead to
\ben
U^2 &= & r^2 - 2 M r - l^2 + b^2 - Q^2,\ a^2 \, V^2 = b^2 \sin^2 \theta, 
\nonumber \\
a\lambda &=& l + b \cos\theta, \,\bar\alpha = 2\frac{b(2Mr+2l^2-Q^2) \sin^2\theta + 2lU^2\cos\theta}{U^2-b^2 \sin^2\theta}.
\een

 Further note that all the metric functions now involve only the four 
parameters, $M,\ b, \ Q, \ l$, indicating removal of the redundant parameter. 
A straight forward calculation would, on dropping overhead bars and replacing 
$b$ by $a$,
take the metric (16) to the form,
\be
ds^2 = \frac{U^2}{\rho^2}\left(dt - P d\varphi\right)^2 - \frac{\sin^2\theta}
{\rho^2}\left(( F+ l^2)d\varphi - a dt\right)^2 \nonumber \\
- \frac{\rho^2}{U^2}dr^2 - \rho^2 d\theta^2 
\ee
where
\be
\rho^2=r^2 +( l+ a \cos\theta)^2, \ F - U^2=2 M r+l^2 - Q^2, \ P- a \sin^2\theta=-2 l \cos\theta, \ F = r^2 + a^2.
\ee
 This is the charged Kerr - NUT solution where $l$ is the NUT parameter 
representing the gravomagnetic monopole charge [3,12] and $Q$ is the electric 
charge. The electromagnetic field $2$-form would be given by
$$
{\bf F} = Q\rho^{-4}\left(r^2 - (l + a\cos\theta)^2 \right){\bf d}r\wedge({\bf d}t - 
P{\bf d}\varphi) + 2Q\rho^{-4}ar\cos\theta\sin\theta{\bf d}\theta\wedge\left((F + l^2){\bf d}\varphi - a{\bf d}t \right).
$$
It reduces to the Kerr - NUT solution [3] when $Q=0$. The general 
electrovac solution we have found is therefore actually the Kerr - NUT 
spacetime with an electric charge. It is asymptotically non flat, and the NUT 
parameter is the measure of asymptotic non flatness. Now when we set $a = 0$, 
it reduces to NUT spacetime with an electric charge. In contrast, the metric 
(16) reduces to the charged black hole with a deficit angle. The metric (18) 
therefore fully encompasses the NUT character of spacetime.

 We have thus found the general solution of the Einstein electrovac equation 
for the stationary axially symmetric spacetime having the integrable metric. 
For vacuum it reduces to the known Kerr - NUT solution. It is the general 
solution for a rotating gravitational dyon, particle which is endowed with 
both gravoelectric 
and gravomagnetic charges. There exists the duality relation between the two 
which is considered in the next section.

\section{Dual and Radiating spacetimes}

{\it (a) Dual spacetime:}
 In the metric (16), the solution is ultimately specified by the 
prescription of the two functions, one of the radial coordinate $r$ and the 
other of the angle coordinate $\lambda$. Let us denote these two functions 
as follows:
$${\cal R}(r) = F - U^2, \,  a^2\Lambda(\lambda) = G + a^2V^2.$$
The Kerr family is characterized by
\be
{\cal R} = 2 M r , \,   \Lambda = 0.
\ee 
On the other hand,
\be
{\cal R} = 0, \,  a^2\Lambda = 2l\lambda
\ee
would characterize the family dual to the Kerr [7]. Here we have considered 
the vacuum case with $Q=0$, and $l=aN$. Eq. (17) 
characterizes Kerr solution when $l=0$ and the dual Kerr solution when $M=0$. 

 The Kerr solution goes over to the dual solution 
under the transformation
\be
 M\rightarrow il, \ r\leftrightarrow ia\lambda
\ee
and the vice-versa. This is the duality relation between the gravoelectric and 
gravomagnetic charges. It takes the field of the one to that of the other.

{\it (b) Radiating spacetime:}
 One of the ways to make a static or stationary spacetime to radiate is to 
transform it into the Eddington retarded coordinates and make mass $M$ 
function of the retarded time. By this method, Kerr solution could be 
transformed into a radiating spacetime having trace free stresses which 
asymptotically conform to the Vaidya radial null flux [10]. The same procedure 
could be applied to turn the Kerr - NUT family (18) radiating. This is what we 
shall demonstrate now.

 Under the transformation,
$$dt\rightarrow dt - \frac{r^2+a^2+l^2}{U^2}dr, \ d\varphi\rightarrow 
d\varphi - \frac{a}{U^2}dr$$
the metric (18) for the case $Q=0$ transforms into the Eddington retarded 
coordinates to read as,
\bea
ds^2&=& \nonumber \frac{U^2-a^2 \sin^2\theta}{\rho^2}dt^2 - 2 \, dt \,dr + 4\frac{a(Mr+l^2)
\sin^2\theta + l U^2 \cos\theta}{\rho^2}dt d\varphi \nonumber \\
&+& 2(a \sin^2\theta-2 l \cos\theta)dr d\varphi - \rho^2d\theta^2 \nonumber \\
&-& \frac{1}{\rho^2} \, \left((r^2+a^2+l^2)^2 \, \sin^2 \theta - (a \sin^2\theta - 2 l \cos\theta)^2
U^2 \right)d\varphi^2,
\eea
where $M=M(t)$ is an arbitrary function of the retarded Eddington time $t$ 
as defined above. This is the Kerr - NUT radiating spacetime with 
trace free stresses which asymptotically go over to the Vaidya null 
radiation flux. When $l=0$, it reduces to the Kerr radiating spacetime [10]. 
The non zero stresses are given in the Appendix A, and it can be easily seen 
that $R = 0$, and they asymptotically reduce to
$$R^r{}_t = -\frac{\dot M}{r^2}, \, R^r{}_{\varphi} = -\frac{\dot M}{r^2}(4l 
\cos\theta - 2a \sin^2\theta)$$
which conform to the Vaidya null radiation [8,9]. When $a = l = 0$, only 
$R^r{}_t$ survives, and it is then the Vaidya radiating star solution. The 
asymptotic behavior is similar when $a$ and/or $l$ are non zero. 

\section{Discussion}
 From the standpoint of physical understanding and application, it is 
important that motion is integrable in the spacetime metric which we wish to 
seek as a solution of the Einstein equation. By integrability of motion, we 
mean separability of HJ and KG equations which respectively govern motion 
of particles and scalar field. Taking this as the general guiding principle 
and an over riding concern, it therefore becomes imperative to tailor the 
metric a priori for integrability. As in [5], we have thus first considered 
separability of HJ and KG equations, which led to the form of the metric as 
given in Eq. (8). Then the non zero Ricci components assume very simple form 
which readily yields to give the general solution given by Eqs (13, 16-19). 
It is the most general solution 
of the Einstein electrovac equation for the integrable axially symmetric 
stationary spacetime. Further it can be transformed to the Kerr - NUT 
solution thereby establishing uniqueness of the Kerr - NUT family for the 
integrable metric. The NUT 
parameter is the cause of asymptotic non flatness. When we impose asymptotic 
flatness, the NUT parameter vanishes and it reduces to the Kerr family. Like 
the Kerr solution, the Kerr - NUT solution could also be made radiating 
by letting the mass parameter be an arbitrary function of the retarded 
Eddington time. The spacetime is non empty but has trace free stresses which 
asymptotically conform to the Vaidya null radiation as is the case for the 
Kerr solution [10].

 The assumption of integrability incorporates existence of regular horizon, 
which is required by separability of KG equation, but not of asymptotic 
flatness.
That is why it leaves room open for asymptotically non flat generalization 
of the Kerr family. It is interesting that this generalization is unique and 
is given by the Kerr - NUT family. Since the Kerr metric is integrable, any 
metric which would contain it would also have to be integrable. For the 
integrable 
metric, the Kerr - NUT family is the most general solution. In other way, we 
have identified integrability of the metric as the uniqueness condition for 
the Kerr - NUT family. If we ask, what is the most general electrovac 
spacetime which includes the Kerr family? The answer is that it is the Kerr - 
NUT family and it is unique. The separability of KG equation can be traded off 
for existence of regular horizon. That is, the requirement of HJ separability 
and regular horizon is equivalent to the metric integrability. In the Kerr 
uniqueness conditions if we trade off asymptotic flatness for HJ separability, 
we arrive at the Kerr - NUT uniqueness conditions. On the other hand HJ 
separability and the vacuum equation imply KG separability [11]. In the the 
asymptotic limit all other metric coefficients go over to the Minkowski form 
but for $g_{t\varphi}$ which is the cause for asymptotic non flatness. Its 
source is the gravomagnetic charge, NUT parameter $l$. As asymptotic 
flatness is the characteristic of the field of the gravoelectric charge (mass), the asymptotic non flatness is similarly the characteristic of the 
gravomagnetic charge (NUT parameter). 

 Our general solution is therefore the most general axially symmetric 
stationary solution for the spacetime having integrable metric. It is thus 
the unique solution for a rotating gravitational dyon. 

 The vacuum spacetime describing a non radiating source would in 
general be stationary and  axially symmetric. Further the metric should be 
integrable if it has to describe a physically meaningful and understandable 
situation. We have obtained the most 
general solution under these physically motivated conditions. It would 
therefore describe the most general source which could in general 
have both gravitational electric and magnetic charge. This is a separate 
matter that it happens to be a transform of the Kerr - NUT solution. Our 
general electrovac solution thus describes the most general particle in 
general relativity. For the integrable spacetime, it is unique 
which implies that there cannot exist any other electro-vacuum solution. 
 
 No sooner gravitational electric and magnetic charges are admitted, the 
question of their duality becomes pertinent. The most remarkable feature of 
the general solution is its invariance under the duality transformation 
$M\leftrightarrow il, \ r\leftrightarrow ia\lambda$ exhibiting duality 
between the gravoelectric charge, mass $M$ and the 
gravomagnetic charge, NUT parameter, $l$. The parameter $a$ plays the 
critical role in this duality transformation, which makes sense only when $a$ 
is non zero. This duality naturally leads to dual of the Kerr solution which 
is also a non trivial vacuum spacetime. This would however not lead to dual 
to the Schwarzschild solution. The dual to the Kerr solution has naturally 
sprung up once the duality relations keeping the metric Eqs (16,17) invariant
 was noticed. This has happened because of the form (17). Else the Kerr - NUT 
solution had been known for over 40 years, yet there had been no 
consideration of spacetime dual to the Kerr solution before our report in 
[7]. In the usual Boyer - Lindquist form given in Eqs (18,19), the duality is 
however not so apparent. 

 It is also to the credit of our method of first seeking the spacetime metric 
to be integrable [5] that the ultimate equations become very simple (see in 
Sec. III) and which readily yield the general solution. This can also be used 
to prove the uniqueness of the Kerr solution when asymptotic flatness is 
imposed. Contrast it with the effort and time spent in proving the uniqueness 
of the Kerr solution (see [13]). We would strongly argue that this could in 
general be a very useful and efficient strategy while seeking physically 
interesting solutions of the Einstein equation. 

 As the Kerr solution describes the gravitational field of a rotating 
gravoelectric monopole, the dual solution 
describes the gravitational field of a rotating gravomagnetic monopole. Even 
though the dual solution is not asymptotically flat, it would be interesting 
to do all that which has been done for the Kerr geometry in the dual Kerr 
geometry so as to gauge the physical effects of asymptotic non flatness. To 
begin with, we would in a separate paper study motion, which is a priori 
integrable, in the Kerr - NUT spacetime with particular reference to the dual 
solution [14].
 
{\bf Acknowledgment:} We thank the anonymous referee for pointing out some 
relevant references. ZT thanks ICTP for a travel grant under its BIPTUN 
program and IUCAA for warm hospitality which made this work possible.


\appendix
\section{}

The non zero components of Ricci tensor for the metric (25) are given by 
\bea
R^t{}_t &=& \frac{-4\,\left( a^2 + 2\,l^2 - 2\,r^2 + 
      4\,a\,l\,\cos \th + 
      a^2\,\cos 2\,\th \right) \,
    {\left( -2\,l\,\cot \th + 
        a\,\sin \th \right) }^2\,{\dot M}}{X^3} \\
R^t{}_{\varphi} &=& \frac{4\,\left( a^2 + 2\,l^2 - 2\,r^2 + 
      4\,a\,l\,\cos \th + 
      a^2\,\cos 2\,\th \right) \,
    {\csc \th}^2\,
    {\left( -2\,l\,\cos \th + 
        a\,{\sin \th}^2 \right) }^3\,{\dot M}}
    {X^3} \\
R^r{}_{\th} &=& \frac{- 2 r \left(-(a^2 - 8 l^2) \cos\th + a (2 l + 6 l \cos 2\th + a \cos 3\th ) \right) \csc\th \dot{M}}{X^2} \\
R^{\th}{}_t &=& \frac{ - 4 r \left(-(a^2 - 8 l^2) \cos\th + a (2 l + 6 l \cos 2\th + a \cos 3\th ) \right) \csc\th \dot{M}}{X^3} \\
R^{\th}{}_{\varphi} &=& \frac{- 16 r( 1 + a \cos\th) \csc\th ( - 2 l \cos\th + a \sin^2 \th)^2 \dot{M}}{X^3} \\
R^{\varphi}{}_t &=& \frac{-4 (a^2 + 2 (l^2 - r^2) + 4 a l \cos\th + a^2 \cos 2\th) \csc^2\th (-2 l \cos\th + a \sin^2 \th) \dot{M}}{X^3} \\
R^{\varphi}{}_{\varphi} &=& \frac{4 (a^2 + 2 (l^2 - r^2) + 4 a l \cos\th + a^2 \cos 2\th)  (-2 l \cos\th + a \sin^2 \th) \dot{M}}{X^3}
\eea

\bea
R^r{}_t &=& \nonumber \frac{-\left( -a^4 +
(12 a^2 + 16 r^2)(r^2 + l^2) +  8 a l(a^2 + 2(l^2 + r^2)) \cos\th +
4 a^2 (5 l^2 + r^2) \cos 2\th + 8 a^3 l \cos 3 \th + a^4 \cos 4\th \right)  \dot{M}}{ X^3} \\
&\hspace{0.2cm}& + \frac{4 r}{\sin^2\th \, X^2} (-2 l \cos\th + a \sin^2 \th)^2 \ddot{M}  \\
R^r{}_{\varphi} &=& \nonumber \frac{2 \left(a^4 + 3 a^2 l^2 + 2 l^4 - 3 a^2 r^2 - 4 l^2 r^2 - 6 r^4 + (4 a l \cos\th  + a^2 \cos 2\th)(a^2 + l^2 - r^2)
\right) (4 l \cos \th - 2 a \sin^2\th)\dot{M}}{ X^3 } \\
&\hspace{0.2cm}& + \frac{2 (- 2 l \cos\th + a \sin^2 \th)^2 (4 l \cos \th - 2 a \sin^2\th) \ddot{M}}{\sin^2\th \, X^2} \\
\eea

where
$$
X = \left(a^2 + 2 l^2 + 2 r^2 + 4 b l \cos\th + b^2 \cos^3 2 \th \right).
$$
Clearly $R = R^t{}_t + R^{\varphi}{}_{\varphi} = 0$.


\begin{references}{}
\bibitem[a]{e-mail:} Email: nkd@iucaa.ernet.in 
\bibitem[b]{e-mail:} Permanently at INP, Ulugbek, Tashkent 702132, Uzbekistan. Email: zafar@suninp.tashkent.su 
\bibitem{1}
A Tomimatsu and H Sato, Phys. Rev. Lett. {\bf 29}, 1344 (1972).
\bibitem{2}
----------------------, Prog. Theo. Phys. {\bf 50}, 59 (1973).
\bibitem{3}
M Demianski and E T Newman, Bull. Acad. Polon. Sci. {\bf 14}, 653 (1966).
\bibitem{4}
B Carter, Comm. Math. Phys. {\bf 10}, 280 (1968).
\bibitem{5}
Z Y Turakulov, Int. J. Mod. Phys. {\bf A4}, 2953 \& 3653 (1989).
\bibitem{6}
J F Plebanski and M Demianski, Ann. Phys. {\bf 98}, 98 (1976).
\bibitem{7} 
Z Y Turakulov and N Dadhich, Mod. Phys. Lett. {\bf A16}, 1959 (2001),
gr-qc/0106042.
\bibitem{8}
P C Vaidya, L K Patel, Phys. Rev. {\bf D7}, 3590 (1973).
\bibitem{9}
P C Vaidya, L K Patel and P V Bhatt, Gen. Relativ. Grav. {\bf 7}, 701 (1976).
\bibitem{10}
M Carmeli and M Kaye, Ann. Phys. {\bf 103}, 97 (1977).
\bibitem{11}
S Bonanos, Comm. Math. Phys. {\bf49}, 53 (1976).
\bibitem{12}
D Lynden-Bell and M Nouri-Zonoz, Rev. Mod. Phys. {\bf 70}, 427 (1998).
\bibitem{13}
S Chandrasekhar, {\it The Mathematical Theory of Black Holes}, 
Clarendon Press, Oxford, 1983.
\bibitem{14}
N Dadhich, S Joshi and Z Y Turakulov, to be submitted.
\end{references}
\end{document}